\newcommand{\ls}[1]     
{\dimen0=\fontdimen6\the=#1\dimen0
 \advance\lineskip.5\fontdimen5\the\lineskip-\dimen0
 \lineskiplimit=.9\lineskip
 \baselineskip=\lineskip     \advance\baselineskip\dimen0
 \normallineskip\lineskip
 \normallineskiplimit\lineskiplimit
 \normalbaselineskip\baselineskip
 \ignorespaces
}

\documentclass[9pt]{extarticle}
\usepackage{spconf,amsmath,graphicx}
\usepackage{xcolor}
\usepackage{multirow}
\usepackage{multicol}
\usepackage{subfigure}
\usepackage{cite}
\usepackage{comment}
\usepackage{hyperref}

\title{Speech Recovery for Real-World Self-powered Intermittent Devices}
\name{Yu-Chen Lin$^{1,3}$, Tsun-An Hsieh$^{3}$, Kuo-Hsuan Hung$^{3}$, Cheng Yu$^{3}$, Harinath Garudadri$^{5}$, Yu Tsao$^{3}$, Tei-Wei Kuo$^{1,2,4}$}
\address{
\small $^1$Department of Computer Science and Information Engineering, National Taiwan University, Taipei, Taiwan\\
\small $^2$NTU High Performance and Scientific Computing Center, National Taiwan University, Taipei, Taiwan\\
\small $^3$Research Center for Information Technology Innovation, Academia Sinica, Taipei, Taiwan\\
\small $^4$Department of Computer Science, City University of Hong Kong, Hong Kong\\
\small $^5$Qualcomm Institute, University of California, San Diego, USA
}

\begin{document}
%
\maketitle
\begin{abstract}

The incompleteness of speech inputs severely degrades the performance of all the related speech signal processing applications.
Although many researches have been proposed to address this issue, they controlled the data missing conditions by simulation with self-defined masking lengths or sizes.
Besides, the masking definitions are different among all these experimental settings.
This paper presents a novel intermittent speech recovery (ISR) system for real-world self-powered intermittent devices.
Three contributive stages: interpolation, enhancement, and combination are applied to the ISR system for speech reconstruction.
The experimental results show that our recovery system increases speech quality by up to 591.7\%, while increasing speech intelligibility by up to 80.5\%.
Most importantly, the proposed ISR system improves the WER scores by up to 52.6\%.
The promising results not only confirm the effectiveness of the reconstruction but also encourage the utilization of these battery-free wearable/IoT devices.


\end{abstract}

\begin{keywords}
internet-of-things, energy harvesting, intermittent systems, speech signal processing, speech recovery
\end{keywords}

\ls{0.9}

\section{Introduction}
\label{sec:intro}

The recent emergence of Internet of things (IoT)/wearable devices has exponentially increased the penetration of IoT applications.
Most of these applications interact with the surrounding environment and communicate with each other by exchanging sensing data (such as temperature, speech signals, or video streams).
More powerful devices such as personal computers (PCs) or servers subsequently realize/construct various complicated applications based on the sensing data received from IoT/wearable devices.
Examples of such advanced applications include image recognition~\cite{CV1}, speech recognition~\cite{ASR1,google_asr,SE2}, and speech enhancement~\cite{SE1,SE2,SE3,SE5,SE6,SE8}.

In the IoT era, wearable devices have begun to outnumber humans; however, most of these devices are powered by batteries, which are often expensive to maintain and can cause serious environmental pollution.
This issue raises the need for environmental energy harvesting to replace battery recharging and thereby alleviate expensive maintenance overheads for wearable devices.
However, the power supplied by ambient sources such as solar, thermal, or vibration energy sources are inherently unstable and sometimes weak~\cite{EHM}.
In contrast to conventional battery-powered systems, intermittent systems relying on energy harvesting typically suffer from insufficient power, thereby resulting in frequent power failures and task interruptions~\cite{NVP2}.
This power insufficiency leads to intermittent sensing data (such as intermittent speech signals) which will be transmitted to PCs or servers for processing.
The incompleteness of the sensing data can severely degrade the performance of all the related applications.
In cross-device speech applications, the intermittent speech results in worse speech quality and inaccuracy of speech recognition.
Many attempts have focused on the incompleteness problem of speech signal processing.
Some works~\cite{ip1,ip2,ip3} imported the concept of inpainting from computer vision into speech audio applications, denoted as audio or speech inpainting.
However, they controlled the data missing conditions by simulation with self-defined masking lengths or sizes.
Besides, the masking definitions are different among all these experimental settings.
Some researches~\cite{plc1,plc2,plc3} have addressed the packet-loss problem in network scenario.
However, the continuous missing samples of the small packet sizes are much fewer than those of power-off duration in intermittent environments.


In the present work, we propose a novel intermittent speech recovery (ISR) system for real-world self-powered intermittent devices
Three contributive stages: interpolation, enhancement, and combination are applied to the ISR system for speech reconstruction.
First, the null segment interpolation initializes the lost areas of the intermittent speech.
Second, the Deep-learning (DL)-based enhancement model using perceptual loss addresses the performance of speech enhancement and recognition.
Finally, the intermittent speech combination stage overcomes the missing-feature problem.
To evaluate the performance enhancement resulting from our approach, we use standardized objective evaluation metrics including the perceptual evaluation of speech quality (PESQ)~\cite{PESQ}, short-time objective intelligibility measure (STOI)~\cite{STOI}, and word error rate (WER)~\cite{WER}.
Experimental results illustrate that our recovery system affords a PESQ-score increase by up to 591.7\% when compared with the original intermittent speech signals, and furthermore, the STOI scores show an increase of up to 80.5\%.
For speech recognition performance, our ISR system also enhances the WER scores by up to 52.6\%.
Even though self-powered devices function with weak energy sources, our ISR system can still maintain the performance of most speech-signal-based applications.
These promising results not only confirm the effectiveness of the reconstruction but also encourage the utilization of these battery-free wearable/IoT devices.

\section{Background and Related Work}
\label{sec:background}

\begin{figure*}[ht]
	\centering
	\includegraphics[width=2\columnwidth]{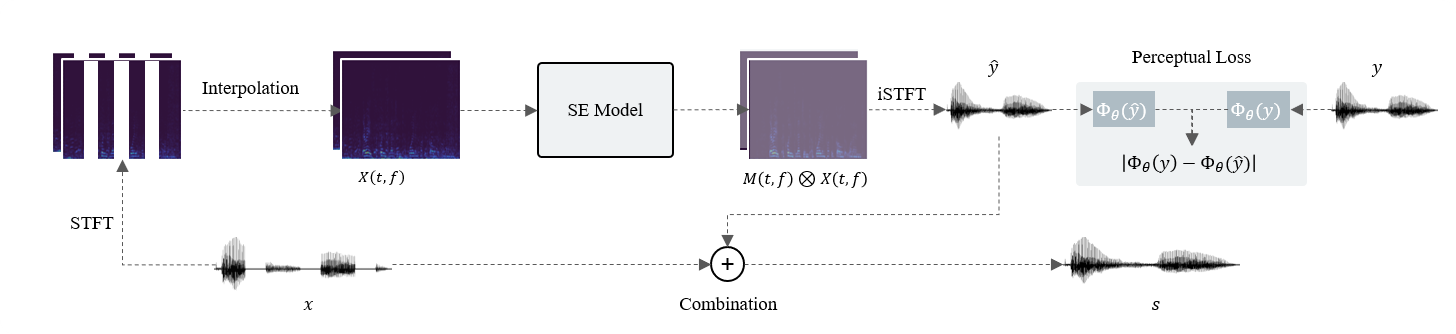}
	\caption{System architecture of the proposed ISR system.}
	\label{fig:network}
\end{figure*}

\begin{figure}[ht]
\centering
\includegraphics[width=0.95\columnwidth]{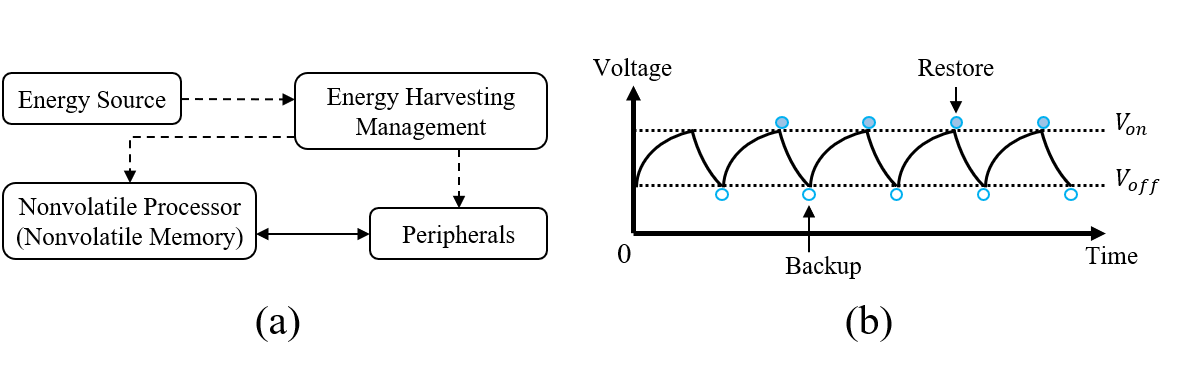}
\vspace{-0.5em}
\caption{(a) self-powered intermittent system (b) general voltage trace of self-powered intermittent systems~\cite{NVP2}.}
\label{fig:EHM}
\end{figure}

A typical self-powered intermittent system consists of an energy harvesting management (EHM), a non-volatile processor (NVP) with non-volatile memory (NVM), and peripherals, as shown in Fig.~\ref{fig:EHM}(a).
The dash-line and the solid-line represent power supplying path and data path, respectively.
The energy source harvests energy from the corresponding ambient source and then transmits the harvested energy to the EHM, which stores energy in its capacitor.
In addition, two voltage thresholds, $V_{on}$ and  $V_{off}$, are the criteria for power supplying.
When the harvested energy is sufficient (i.e., the capacitor voltage equals $V_{on}$), the EHM supplies power to the system.
When the NVP is activated, it uses the previous backup memory to restore program states and subsequently continues executing programs, operating peripherals, and backing up the program state.
The NVP may support a periodic or on-demand backup mechanism to address the power failure problem~\cite{backup}.
When a power failure occurs (i.e., the capacitor voltage equals $V_{off}$), the NVP suspends operations until sufficient energy is harvested and subsequently resumes instantly from the last operation point after the power is restored.
A general voltage trace of self-powered intermittent systems is shown in Fig.~\ref{fig:EHM}(b).
Therefore, a intermittent system with an energy-harvesting unit can operate intermittently over long duration without any external power source.

Numerous studies have been conducted to exploit intermittent systems (NVM and NVP hardware) for DL-based applications.
Many such attempts have focused on the inference phase for resource-constrained devices, such as achieving in-memory computing~\cite{i1}, while a few attempts have been made to cope with the problems in the training phase~\cite{t2}.
Some studies have even implemented DL-based algorithms on NVP devices~\cite{NVP1}.
However, most DL algorithms require a large amount of memory and computation, and therefore, servers are used to handle most of the computational workload of DL algorithms.
The use of intermittent non-volatile hardware for either inference or training results in performance degradation and an increase in response latency.
That is, the intermittent systems play a negative role in the DL-based IoT applications.
In addition, the low-power features of intermittent systems are suitable for IoT/wearable devices.
As a result, in DL-based IoT applications, an intermittent system should serve as a sensor node collecting environmental information, while the DL algorithm serves as a remote server processing the sensing data.

\section{The ISR System}
\label{sec:method}

The system architecture of ISR is shown in Fig.~\ref{fig:network}.
The ISR system consists of three stages: interpolation, enhancement, and combination.
We detail these three stages in Sections~\ref{sec:3.1} to \ref{sec:3.3}.

\subsection{Null segment interpolation}
\label{sec:3.1}

The intermittent speech signal received by the ISR system contains multiple null segments with time stamps.
These null segments are difficult to recover without the assumption of certain values.
Therefore, we used the time stamps to interpolate these null segments in a linear manner.
In our approach, we interpolate the intermittent speech in the frequency domain rather than the waveform domain for the following reasons:
Interpolation in the waveform domain for intermittent speech signals is empirically undesirable because the reconstructed waveform signal simply carries the information of the original signal.
The speech signal is constructed by the fundamental frequency and its harmonics, which are more observable when using time–frequency analysis or a log-powered spectrum.
When we map the intermittent signal to the frequency domain, we can perform interpolation that serves in adequately reconstructing the frequency components of speech in the null segments.
For one null segment that starts at $t_1$ and ends at $t_2$, we calculate the weighting ratio of each frame in this null segment using Eq.~\eqref{eq:r}:
\begin{align}
\label{eq:r}
    r(t) = \frac{t - (t_1 - 1)}{(t_2 + 1) - (t_1 - 1)}
\end{align}
where $t_1-1$ denotes the index of the frame immediately before the null segment and $t_2+1$ denotes the index of the frame immediately after the null segment.
After calculating ratio $r(t)$, we interpolate values $X(t,f)$ using Eq.~\ref{eq:X}:
\begin{equation}
\label{eq:X}
    X(t,f) = (1 - r(t))X(t_1-1,f)+r(t)X(t_2+1,f)
\end{equation}
where $X(t,f)$ denotes the frame vector of $t$ in the frequency domain.
In addition, we need to consider two cases: 1) $t_{1}$ is the first frame and 2) $t_{2}$ is the last frame of the signal.
In these edge cases, we calculated $X(t,f)$ with zero vector.
Consequently, we can obtain a processed intermittent speech with initialization in all the null segments, as shown in the left side of Fig.~\ref{fig:network}.

\subsection{DL-based enhancement model}
\label{sec:3.2}

Once the null segments are coarsely interpolated, a DL-based enhancement model is used for further refinement.
We use a complex U-Net~\cite{choi2018phase} architecture, which is composed of specialized convolutional layers that follow the rule of the addition and multiplication of complex numbers.
The main reason for applying a complex-based model is to preserve phase information in the procedure of intermittent speech signal recovery.
As compared to magnitude-spectrum-mapping-based models that use the unprocessed phase spectrum to convert the frequency-domain signals back to the time domain by inverse short-time Fourier transform (iSTFT), the complex U-Net directly uses the complex spectrum as its input and output.
Therefore, phase information is well considered during the enhancement.
In addition, the enhancement model generates complex masks, which are then element-wise multiplied by each interpolated spectrum, to obtain the recovered spectrum.

Several studies~\cite{germain2018speech,tahsieh} have demonstrated that perceptual losses (also known as feature losses) are effective in improving the perceptual quality in speech enhancement tasks.
The success of the approach is attributed to the fact that the perceptual losses measure the distances in some latent domains invariant to short-time shifts.
Perceptual loss is defined as the average of the mean absolute error (MAE) between features, extracted by a loss model $\Phi$ parameterized by $\theta$, and calculated by Eq.~\eqref{eq:loss}:
\begin{align}
    \label{eq:loss}
    \mathcal{L}(y, \hat{y})=\frac{1}{CL}\sum_{c=1}^{C}\sum_{l=1}^{L}|\Phi_\theta(y)_{c, l} - \Phi_\theta(\hat{y})_{c, l}|
\end{align}
where $C$ denotes the number of channels; $L$ represents the length of the feature; $y$ and $\hat{y}$ denote ground truth and enhanced speech signal, respectively.
Here, as the recovered output speech signals are expected to be used in speech applications, we choose a self-supervised encoder~\cite{schneider2019wav2vec}, which renders representative speaker and phonetic information, for automatic speech recognition (ASR) as our loss function.
This approach enables the enhancement system to achieve a better performance in terms of both speech quality and recognition accuracy relative to common signal-level losses such as MAE and the mean squared error (MSE).

\subsection{Combination}
\label{sec:3.3}

After the processing of the interpolated intermittent speech signals, the enhanced speech signals are generated by the DL-based enhancement model.
Because of the features of regression models, the enhancement model not only processes signals in the null segments but also changes the values in the originally existing segments.
However, the adjustment of existing values may result in missing features, i.e., features such as phonemes, frequency, and amplitude may be influenced.
Therefore, we combined the original intermittent speech $x$ with the enhanced speech $\hat{y}$ through an indicator-like combination to generate the final recovery speech $s$.
In this study, we simply apply the combination in time-domain (i.e., waveforms) for convenience.
Accordingly, recovery speech $s$ is calculated using Eq.~\eqref{eq:s}:
\begin{align}
\label{eq:s}
s(t)=
\begin{cases}
\hat{y}(t), & {\rm if\ }t{\rm \ in\ null\ segments} \\
x(t), & {\rm else}
\end{cases}
\end{align}
where $t$ denotes the time index of the waveform.
In the equation, we directly used original values $x(t)$ for the existing segments while using enhanced speech value $\hat{y}(t)$ from the U-Net-based model for the null segments.
With this approach, the intermittent speech also carries the original features when being recovered in the null segments.

\section{Performance Evaluation}
\label{sec:evaluation}

\subsection{Experimental setup}
\label{sec:4.1}

Tab.~\ref{tab:spec} details the specifications of the experimental apparatuses.
For the intermittent microphone device, the backup and restore voltage thresholds are 2.3 V and 2.8 V respectively.
We used the MAX98089 Low-Power, Stereo Audio Codec with FlexSound Technology\footnote{https://datasheets.maximintegrated.com/en/ds/MAX98089.pdf} as our microphone sensor, for which the energy consumption for recording was 5.6 mW.
For the EHM and power supply, the capacitance in the EHM was set to 200 µF, the energy source for training ranged from 1.5 to 5.5 mW (in steps of 0.25 mW), and the energy source for testing ranged from 2.0 to 5.0 mW (in steps of 1.0 mW).
In addition, the scenarios of training and testing were mismatched for reality, that is, the training energy source excluded the 2.0, 3.0, 4.0, and 5.0 mW cases.
The periods of power-on and power-off can be calculated by the following capacitor potential energy formula~\cite{jackson}:
\begin{equation}
E = P \times T = \frac{C \times V^2}{2}
\end{equation}
where $E$ is the amount of harvested energy; $P$ is the difference between energy the power of energy source and the recording energy consumption; $C$ is the capacitance; and $V^2$ is the difference between $V_{on}^2$ and $V_{off}^2$.
The on and off periods of testing scenarios are listed in Tab.~\ref{tab:spec} and in units of milliseconds (ms).

\begin{table}[h]
    \centering
    \caption{Specifications of the experimental platform}
	\label{tab:spec}
	\begin{tabular}{|p{4cm}|p{3cm}|}
	\hline
	\multicolumn{2}{|c|}{Intermittent Microphone Device} \\
	\hline
    Restore threshold ($V_{on}$) & 2.8 V \\
	Backup threshold ($V_{off}$) & 2.3 V \\
	Recording energy consumption & 5.6 mW \\
	\hline
	\hline
	\multicolumn{2}{|c|}{EHM \& Power Supply} \\
	\hline
	Capacitance & 200 uF \\
	Training energy source & 1.5 to 5.5 mW \\
	& (with a step of 0.25) \\
	Testing energy source & 2.0 to 5.0 mW \\
	& (with a step of 1.0) \\
	\hline
	\end{tabular}
\end{table}

\begin{table*}[ht]
	\footnotesize
	\caption{
	Detailed PESQ, STOI, and WER scores for the intermittent, interpolated, our proposed three-stage ISR with MSE as the objective function, and the proposed three-stage ISR with perceptual loss.
	Each score is an average score of all 824 testing utterances.
	The score improvement are represented in the percentage from the intermittent to the processed.
	The average WER score of clean speech signals is 0.21.
	}
	\centering 
	\label{tab:res}
	\begin{tabular}{| c | cc | ccc | ccc | ccc | ccc |}
	    
        \hline
		 & \multicolumn{2}{|c|}{Period (ms)} & \multicolumn{3}{|c|}{Intermittent} & \multicolumn{3}{|c|}{Interpolated} & \multicolumn{3}{|c|}{ISR+MSE} &
		\multicolumn{3}{|c|}{ISR+PL} \\
        \hline
		 Energy & \multirow{2}{*}{On} & \multirow{2}{*}{Off} & \multirow{2}{*}{PESQ} & \multirow{2}{*}{STOI} & \multirow{2}{*}{WER} & \multirow{2}{*}{PESQ} & \multirow{2}{*}{STOI} & \multirow{2}{*}{WER} & \multirow{2}{*}{PESQ} & \multirow{2}{*}{STOI} & \multirow{2}{*}{WER} & \multirow{2}{*}{PESQ} & \multirow{2}{*}{STOI} & \multirow{2}{*}{WER}\\ 
		 source & & & & & & & & & & & & & &\\

		\hline
		
		\multirow{2}{*}{2.0} & \multirow{2}{*}{71} & \multirow{2}{*}{128} & 0.24 & 0.41 & 0.99 & 1.30 & 0.53 & 0.99 & 1.01 & 0.51 & 0.97 & 1.66 & 0.74 & 0.86 \\
		& & & - & - & - & 441.7\% & 29.3\% & 0.0\% & 320.8 & 24.4\% & 2.0\% & 591.7\% & 80.5\% & 13.1\% \\
		\hline
		
		\multirow{2}{*}{3.0} & \multirow{2}{*}{98} & \multirow{2}{*}{85} & 0.51 & 0.56 & 0.96 & 1.58 & 0.66 & 0.95 & 1.28 & 0.67 & 0.91 & 2.11 & 0.84 & 0.59 \\
        & & & - & - & - & 209.8\% & 17.9\% & 1.0\% & 151.0\% & 19.6\% & 5.2\% & 313.7\% & 50.0\% & 38.5\% \\
		\hline
		
		\multirow{2}{*}{4.0} & \multirow{2}{*}{159} & \multirow{2}{*}{64} & 0.98 & 0.74 & 0.76 & 1.96 & 0.80 & 0.67 & 1.86 & 0.82 & 0.61 & 2.59 & 0.92 & 0.36 \\
		& & & - & - & - & 100.0\% & 8.1\% & 11.8\% & 89.8\% & 10.8\% & 19.7\% & 164.3\% & 24.3\% & 52.6\% \\
		\hline
		
		\multirow{2}{*}{5.0} & \multirow{2}{*}{425} & \multirow{2}{*}{51} & 1.94 & 0.90 & 0.36 & 2.61 & 0.92 & 0.34 & 2.75 & 0.93 & 0.35 & 3.23 & 0.97 & 0.26 \\
		& & & - & - & - & 34.5\% & 2.2\% & 5.6\% & 41.8\% & 3.3\% & 2.8\% & 66.5\% & 7.8\% & 27.8\%\\
		\hline
		
	\end{tabular}
\end{table*}

In our experiments, we used the VCTK-DEMAND corpus~\cite{VCTK}, a 16KHz corpus with fixed training and testing data.
For the training set, we used 11,572 utterances from the clean training set.
We corrupted these utterances with 13 different energy sources to generate 11,572 $\times$ 13 $=$ 150,436 training utterances.
For the testing set, we used 824 utterances, which were different from the utterances used in the training set, from the testing set of the VCTK-DEMAND corpus.
These utterances were corrupted by 4 different energy sources to generate 824 $\times$ 4 $=$ 3,296 testing utterances.
We compared our recovered speech signals with the intermittent and interpolated speech signals.
To verify the effect of perceptual loss, an ISR model with trained with MSE (a commonly used loss for regression tasks) was also tested for comparison.
To evaluate the model performance, we used three standardized objective evaluation metrics: PESQ~\cite{PESQ}, STOI~\cite{STOI}, and WER~\cite{WER}. 
PESQ indicates the speech quality, and its score ranges from -0.5 to 4.5.
STOI is a well-known objective metric for measuring perceptual speech intelligibility, with scores ranging from 0 to 1.
The WER score indicated the ASR results and was calculated with Levenshtein distance~\cite{WER} using a pre-trained ASR system provided by Google~\cite{google_asr}.

\subsection{Experimental results}

The experimental results are listed in Tab.~\ref{tab:res}.
Please note that the average WER score of clean speech signals is 0.21.
From the table, we first observe that the null segment interpolation can directly and effectively improve the speech quality and intelligibility of the intermittent speech.
The PESQ scores exhibit an improvement between 34.5\% and 441.7\%, while the STOI scores improve by 2.2\% to 29.3\%.
However, the WER scores only marginally improve by 0.0\% to 11.8\%.
The main reason is that the null segment interpolation only takes few features (such as frequency and power) of intermittent speech signals in consideration.
Although the improvement in WER scores using null segment interpolation is marginal, the initialization in null segments is still important for the next enhancement stage.
More specifically, most of the DL-based models, including the strategies mentioned in Sec.~\ref{sec:intro} and our second-stage enhancement model, are not trainable without the initialization.
The main reason is that choosing an appropriate starting point is important for training DL-based algorithms.

In addition, upon applying our proposed ISR with perceptual loss (ISR+PL), we find that the PESQ scores further improve by 32.0\% to 150.0\%, while the STOI scores additionally improve by 5.6\% to 51.2\%.
Meanwhile, the WER scores also additionally improve by 13.1\% to 48.8\% as compared to the interpolated speech signals.
To have a better understanding of the improvement, we compared our recovery model with another ISR based on MSE (ISR+MSE), a common signal-level loss.
It is obvious that ISR+MSE deteriorate the speech quality and intelligibility of the interpolated speech signals.
The PESQ and STOI scores respectively suffered a severe degradation by 120.9\% and 4.9\% under 2.0 mW energy source.
Although the WER scores and some SE performances of the greater energy conditions additionally improve, the improvements are marginal.
The reason is that MSE is not strongly correlated to perceptual evaluation metrics, minimizing MSE does not necessarily leads to high PESQ/STOI and low WER.
On the other hand, our ISR+PL model enhances the WER scores by 13.1\% to 52.6\% when compared with the counterpart scores of intermittent speech.
The main reason for the improvements is that the perceptual loss has not only frequency and power features but representative speaker and phonetic information during training, as mention in Sec.~\ref{sec:3.2}.
These results suggest that our proposed ISR+PL model not only improves the speech quality and intelligibility but also the accuracy of speech recognition.

\begin{figure}[ht]
    \centering
    
    \subfigure[clean waveform]{
        \includegraphics[width=0.47\linewidth]{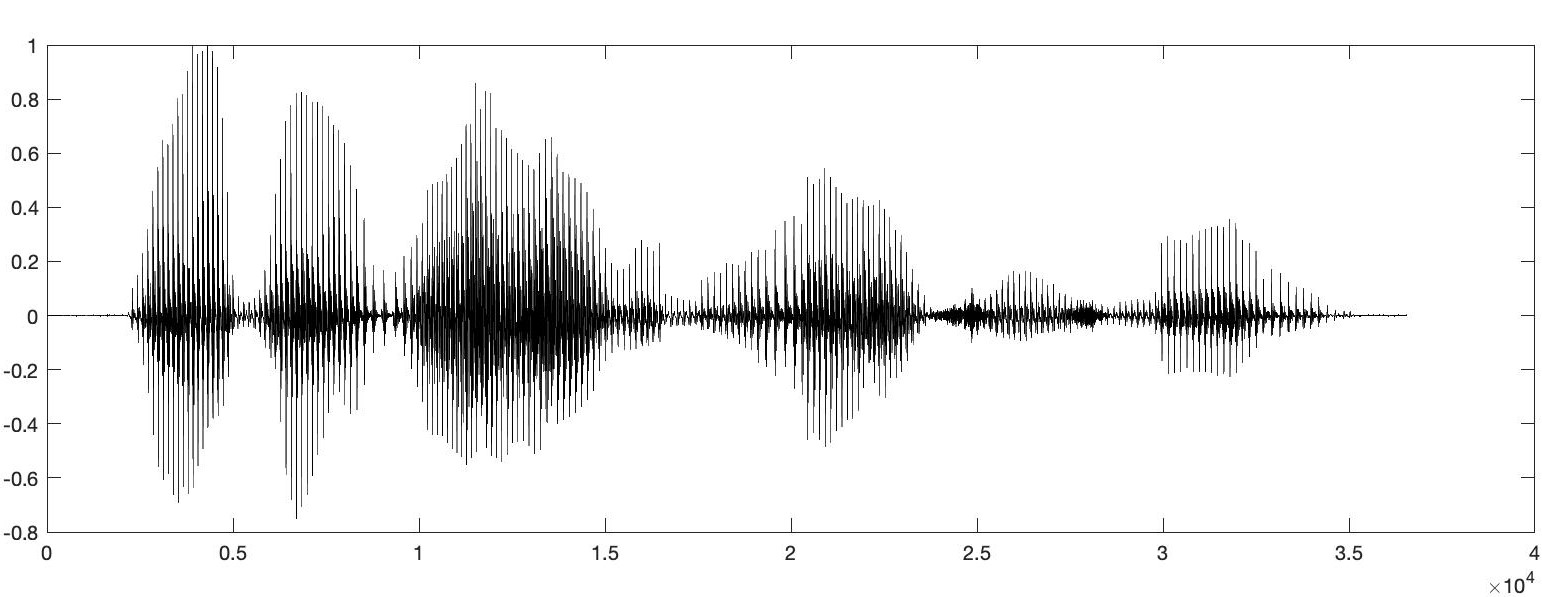}
    }
    \subfigure[clean spectrogram]{
        \includegraphics[width=0.47\linewidth]{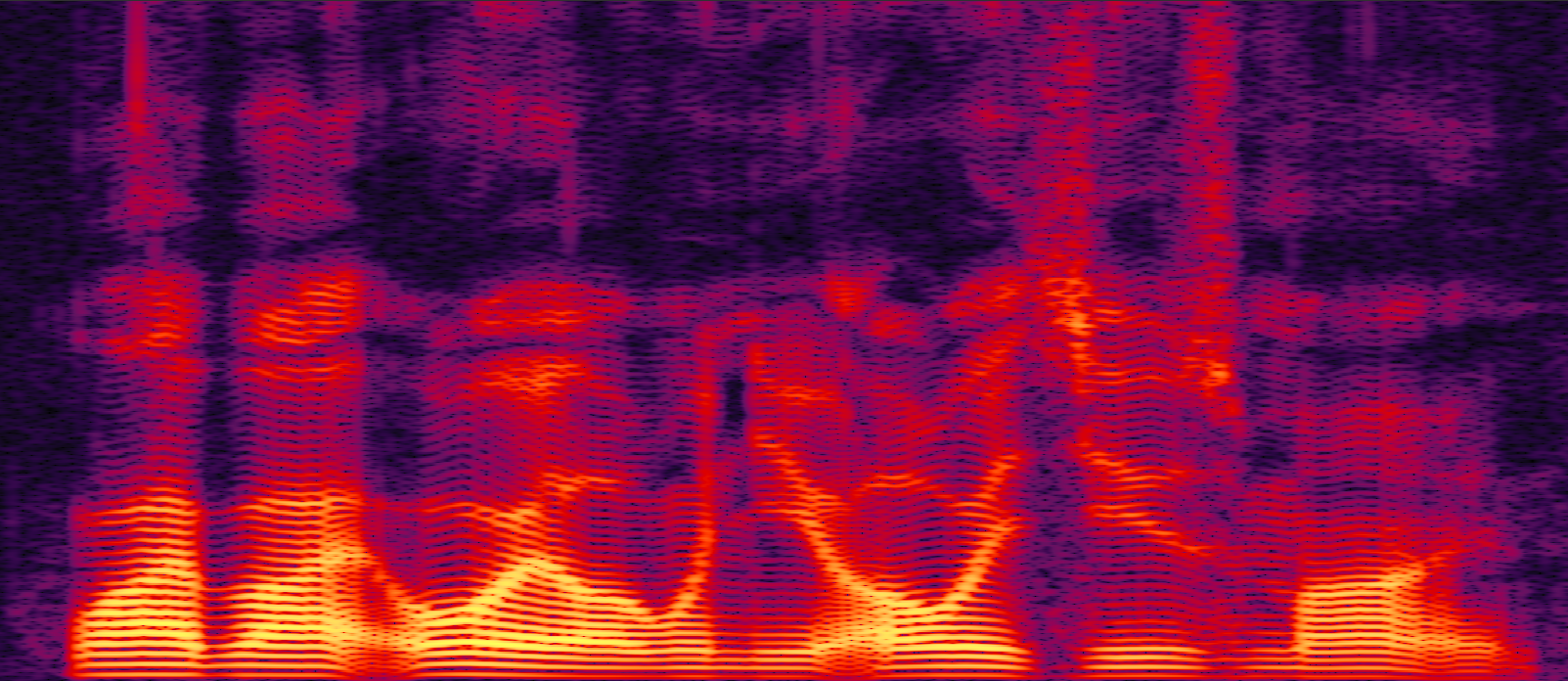}
    }
    \subfigure[intermittent waveform]{
        \includegraphics[width=0.47\linewidth]{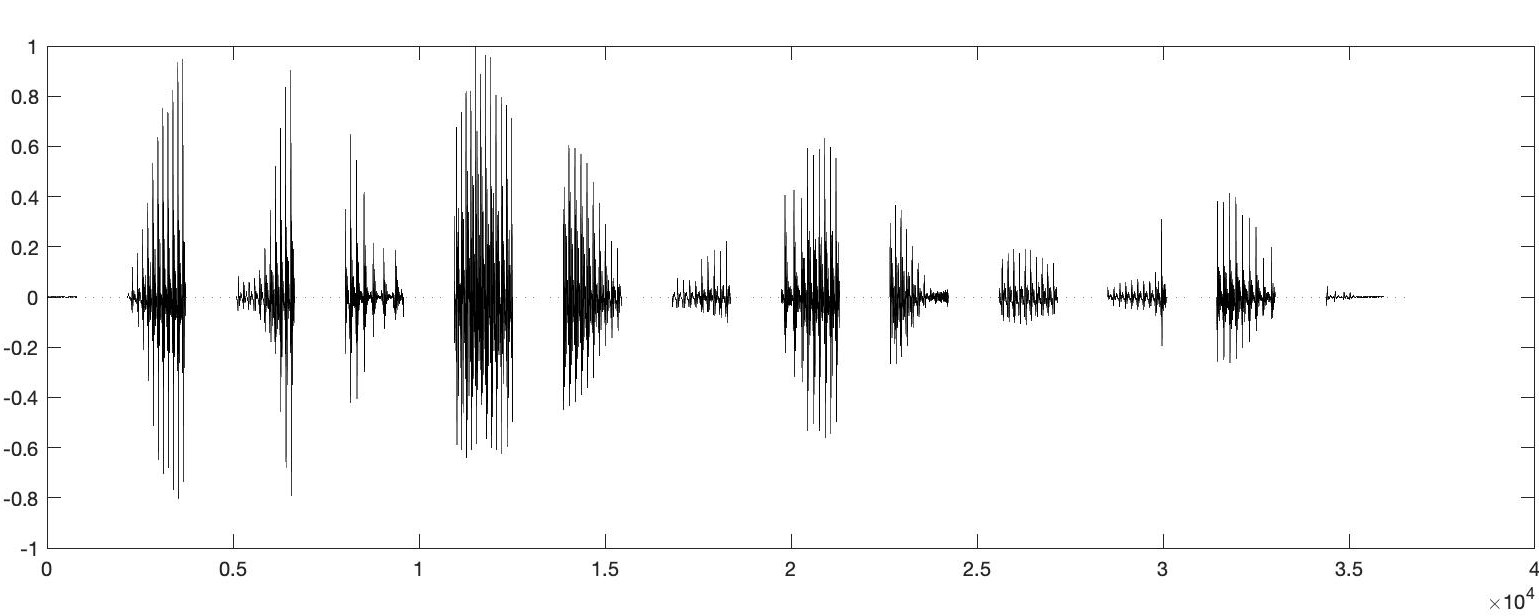}
    }
    \subfigure[intermittent spectrogram]{
        \includegraphics[width=0.47\linewidth]{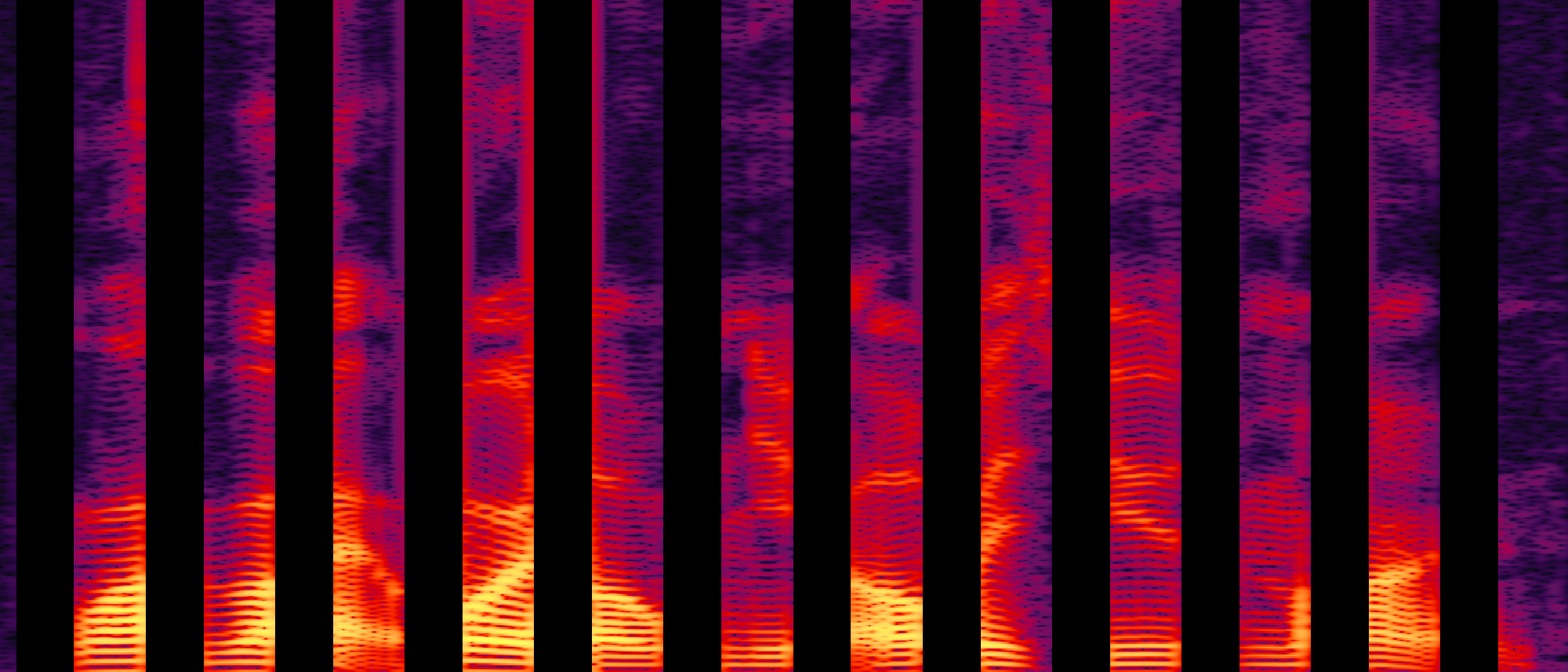}
    }
    \subfigure[recovered waveform]{
        \includegraphics[width=0.47\linewidth]{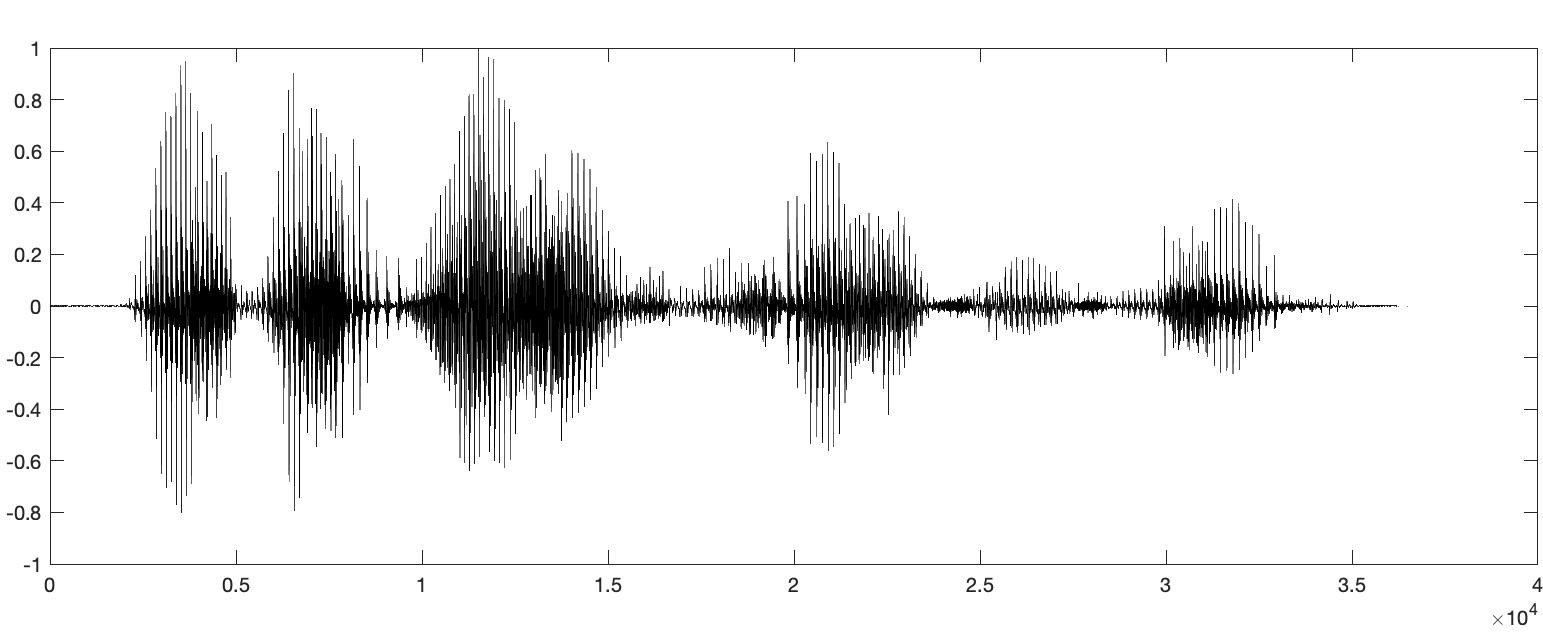}
    }
    \subfigure[recovered spectrogram]{
        \includegraphics[width=0.47\linewidth]{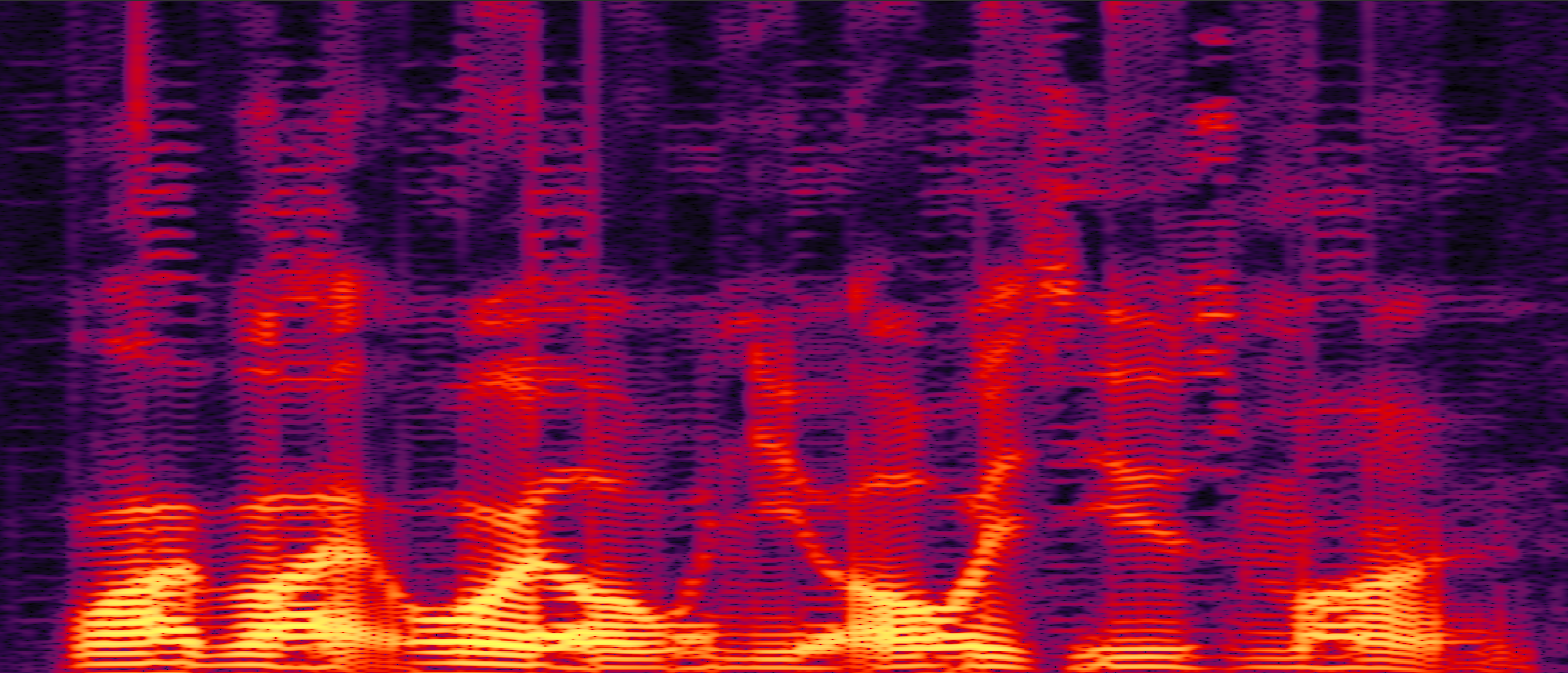}
    }
    \vspace{-0.5em}
    \caption{
    Waveforms and spectrograms of an example utterance: (a) and (b) clean; (c) and (d) intermittent under 3.0 mW energy source; (e) and (f) recovered by the ISR+PL.
    }
    \label{fig:example}
    \vspace{-0.5em}
\end{figure}

We also observed that the improvement decreased as the source energy changed from a low-energy source to a high-energy source in speech quality and intelligibility metrics.
A possible reason for this result is that the poor energy environment of a weak energy source results in the self-powered device being out of power.
As a result, there is still room for further improvement.
However, the intermittent speech signals due to a strong energy source originally perform well across all three metrics, and the improvements are less significant than the counterpart ones in the case of the weak energy source.
Finally, Fig.~\ref{fig:example} shows the speech waveforms and spectrograms of clean, intermittent, and recovered speech, respectively.
We can observe from Fig.~\ref{fig:example} (c) and (d) that there are many null segments in the intermittent speech.
These null segments degrade the speech quality and intelligibility and increase the WER for speech recognition.
Our proposed three-stage ISR with perceptual loss can reconstruct some effective speech signals in these loss areas of the intermittent waveform (Fig.~\ref{fig:example} (e)) and spectrogram (Fig.~\ref{fig:example} (f)).
The example waveforms are available in the file sharing site\footnote{\href {https://github.com/dwadelin/ISR\_Examples}{https://github.com/dwadelin/ISR\_Examples}}.

\section{Conclusions}
\label{sec:conclusion}

In this study, we have proposed an intermittent speech recovery (ISR) system for real-world self-powered intermittent devices.
Three contributive stages: interpolation, enhancement, and combination are applied to the ISR system for speech reconstruction.
First, the null segment interpolation initializes the non-value areas of the intermittent speech signals.
Second, the DL-based enhancement model with perceptual loss addresses the performance of speech quality, intelligibility, and accuracy of recognition.
Finally, the intermittent speech combination overcomes the missing-feature problem.
The experimental results show that our ISR system affords an increase of up to 591.7\% (from 0.24 to 1.66) on PESQ scores when compared with case of the intermittent speech signals, and further, the STOI score increases up to 80.5\% (from 0.41 to 0.74).
Most importantly, our ISR system also enhances the WER scores by up to 52.6\% (from 0.76 to 0.36), as compared to original intermittent speech.
Even though self-powered devices function with weak energy sources, our ISR system can still maintain the performance of most speech-signal-based applications.
These promising results suggest that even though self-powered microphone devices function with weak energy sources, our ISR system can still maintain the performance of most speech-signal-based applications.

\bibliographystyle{IEEEbib}
\bibliography{refs}

\end{document}